\documentclass[a4paper,12pt, reqno]{article}

\usepackage{amssymb,amsmath,setspace,graphicx,mathrsfs, float,slashed, mathrsfs, amsfonts, cite}

\usepackage[utf8]{inputenc}

\def\be{\begin{equation}}
\def\ee{\end{equation}}

\def\bea{\begin{eqnarray}}
\def\eea{\end{eqnarray}}
\def\rks{$R_{K^*}$}
\def\rk{$R_{K}$}
\def\lq{Leptoquark~}
\def\lqs{Leptoquarks~}
\def\gm{$\left(g-2 \right)_\mu$}
\def\da{\Delta^{2/3}}
\def\db{\Delta^{5/3}}

\def\lqa{\Delta^{5/3}}
\def\lqb{\Delta^{2/3}}

\begin{document}

\begin{center}
	\LARGE{Discrepancies in simultaneous explanation of Flavor Anomalies and IceCube PeV Events using Leptoquarks} \\ \vspace{.5cm}
	\large{Bhavesh Chauhan$^{a,b,\#}$, Bharti Kindra$^{a,b,\star}$, Ashish Narang$^{a,b,\dagger}$} \\ \vspace{.5cm}
	\small{$\quad^a$ Physical Research Laboratory, Ahmedabad, India. \\
		$\quad^b$ Indian Institute of Technology, Gandhinagar, India. \\
		$\quad^\#$bhavesh@prl.res.in $\quad^\star$bharti@prl.res.in $\quad^\dagger$ashish@prl.res.in
	}
\end{center}

\textbf{Abstract:}  Leptoquarks have been suggested to solve a variety of discrepancies between the expected and observed phenomenon. In this paper, we show that the scalar doublet Leptoquark with Hypercharge 7/6 can simultaneously explain the recent measurement of $R_{K}$, $R_{K^*}$, the excess in anomalous magnetic moment of muon, and the observed excess in IceCube HESE data. For appropriate choice of couplings, the flavor anomalies are generated at one-loop level and IceCube data is explained via resonant production of the Leptoquark. Several constraints from LHC searches are imposed on the model parameter space. \\


\section{Introduction}

Leptoquarks are solution to the problem of matter unification which appear naturally in many theories beyond the Standard Model (SM). For example, scalar quarks in R-Parity Violating Supersymmetry (RPV) have \lq like Yukawa couplings \cite{farhi} whereas vector \lqs arise in Grand Unification Theories (GUT) based on $SU(5)$ and $SO(10)$ \cite{georgi1, georgi2, cox}. The unique feature of \lqs is that they couple simultaneously to Standard Model (SM) quarks and leptons, thus providing ample testing grounds and applications to variety of discrepancies between theory and experiments. \\

The latest measurement of \rks~and \rk~by LHCb, has pointed towards $\approx 2.5\sigma$ deviation from the standard model \cite{rk, rkstar}. These are clear hints of Lepton Flavor Universality (LFU) violation which can be explained in a wide variety of frameworks including, but not limited to, Leptoquarks \cite{HillerLQ,amico,becirevic,hiller,volkas,crivellin3},  RPV \cite{rpv1,rpv2,rpv3, bhupal2}, E6 \cite{e6}, flavor violating Z$^\prime$ \cite{z1,z2,z3,z4,z5,z6,z7,z8,z9,crivellin2,ghoshnew,bonilla,Cline} etc. In the past, \lqs have been used to explain the anomalous magnetic moment of muon \cite{g22,g23,g24,g25,g26,g27}, flavor anomalies \cite{HillerLQ,amico,becirevic,hiller,volkas,crivellin3}, and IceCube PeV events \cite{lq0,iclq1,iclq2,iclq3,iclq4,iclq5,bhupalice} independently. However, simultaneous explanation of all the three observations has not been possible due to the different range of \lq masses required to solve the individual problems. In this work, we show that a scalar \lq of mass close to $1$ TeV can explain the aforementioned discrepancies. However, such an explanation would be extremely unfavoured by LHC data. While the particular results are model dependent, one can make a qualitative predictions about a more general model. \\

In Section \ref{model} we describe the model of \lq and motivate the texture of the coupling matrices that has been used in this paper. In Section \ref{gminus2} we explain the excess in \gm~using this model. In Section \ref{flavor} we explain the recent measurement of \rk~and \rks~within our framework followed by the explanation for IceCube High Energy Starting Events (HESE) in Section \ref{IC}. In Section \ref{results} we discuss the results of this analysis and obtain the parameter space for simultaneous explanation. In the next section, we do the LHC analysis for the benchmark point and obtain the constraints. In the end, we conclude with some model-dependent and model-independent statements. 

\section{Model Description}\label{model}

In this paper, we consider the scalar \lq $\Delta  = \left(\mathbf{3,~2,}~7/6\right)$ whose interactions with the SM fields is given as \cite{fajfer},
\begin{equation}
\label{lagshort}
\mathcal{L}_{\Delta} \ni  - (y_L)_{ij} \bar{u}^i_R \mathbf{\Delta}_a \varepsilon^{ab} (L_L)_b^j  + (y_R)_{ij} \bar{Q}_{L}^{i~a} \mathbf{\Delta}_a \mathit{l_R}^j + \text{h.c.}
\end{equation}
where $y_{L(R)}$ are the Yukawa-like couplings of the Leptoquark. For simplicity, we have assumed the couplings to be real. We have not shown the kinetic and Higgs interactions for brevity, however they are relevant for the discussion that follows. We refer the reader to reference \cite{fajfer} for a comprehensive analysis. We can rewrite \eqref{lagshort} in terms of the mass eigenstates $\db$ and $\da$, where the superscript denotes electric charge. In terms of these states, the Lagrangian \eqref{lagshort} is written as, 
\begin{align}
\label{lag}
\mathcal{L}_{\Delta} \ni& (V y_R )_{ij} \bar{u}_i P_R l_j \db  - (y_L)_{ij} \bar{u}_i P_L l_j \db  \\ 
&+ (y_R)_{ij} \bar{d}_i P_R l_j \da + (y_L U)_{ij} \bar{u}_{i}P_L \nu_j \da + \text{\emph{h.c}.}
\end{align}
where V and U are the CKM and PMNS matrices respectively. In common literature \cite{fajfer}, this model is also known as $\mathbf{R_2} $. \\

\noindent The observed negligible branching ratios of the flavor violating decays of leptons (for example,  $\tau \rightarrow \mu \gamma$ and $\mu \rightarrow e \gamma$) put stringent constraints on the inter-generation couplings of the Leptoquark. For all practical purposes, this implies that 
\be y_{L(R)}^{qe} = y_{L(R)}^{q\tau} = 0 \quad \forall q. \ee
It has been argued in previous works that the this Leptoquark model results in \rk$\approx 1$, and \rks$\approx 1$ because of the tree level contribution to $b \rightarrow s \mu \mu $  \cite{hiller}. This clearly contradicts the recent measurements by LHCb. It was pointed out in \cite{becirevic} that, if one assumes 
 \be y_R^{s\mu} = 0 \quad \text{or} \quad y_R^{b\mu} = 0, \ee 
then the tree level contribution is negligible and the leading contribution of comes from a one-loop process. It will be shown in Section \ref{flavor} that this results in \rk$<1$, and \rks$<1$ which is in agreement with the latest experiments. We chose the former solution as it is also favoured by \gm. As mentioned in \cite{becirevic}, non-zero $y_L^{c\mu}$ results in tree level contribution to $b \rightarrow c l \bar{\nu_l} $ which contradicts the observed $R(D)$ and $R(D^\star)$. Hence, we also assume that 
\be y_L^{c\mu} = 0.\ee
In order to avoid undesired contribution to other rare decays of the B meson, such as $b\rightarrow d \mathit{l^+l^-}$, we assume that \be y_{R}^{d\mu} \approx 0. \ee  With these constraints, the coupling matrices are,
\be 
\label{coup}
\quad y_L                                                                                                                                                                                                                                                                                = \begin{pmatrix}
	0 & y_{L}^{u\mu} & 0 \\
	0 & 0 & 0 \\
	0 & y_{L}^{t\mu} & 0
\end{pmatrix} \ ,\quad y_R                                                                                                                                                                                                                                                                                = \begin{pmatrix}
	0 & 0 & 0 \\
	0 & 0 & 0 \\
	0 & y_{R}^{b\mu} & 0
\end{pmatrix} . \ee 
For brevity, we will use $  y_{L}^{u\mu}  = \lambda_1,  ~y_{L}^{t\mu} = \lambda_2, \text{and}  ~y_{R}^{b\mu} = \lambda_3 $ for the remainder of this paper. We will also use $M_1$ ($M_2$) to denote the mass of $\db$ ($\da$). \\

\noindent In subsequent sections, it will be pointed out that the LHC constraints limit $M_1 \geq 1100$ GeV. For our analysis, we take the lower limit and generate constraints on the remaining parameters. If future searches increase the lower limit considerably, the expressions will change accordingly. Having said that, there are only four free parameters in our model 
\be \{ M_2, \lambda_1, \lambda_2, \lambda_3 \}. \ee
In the subsequent sections, we investigate various constraints on the model parameters coming from \gm, flavor anomalies, IceCube data, and LHC. 
\section{$\mathbf{\left(g-2 \right)_\mu}$}\label{gminus2}

The experimentally measured value of the anomalous magnetic moment of muon is slightly larger than the prediction from the Standard Model. This discrepancy has been attributed to a variety of new physics scenarios \cite{g21,g22,g23,g24, Belanger}. At present, the difference is  \cite{pdg}, 
  \be\delta a_\mu = a_\mu^{EXP} - a_{\mu}^{SM} = (2.8 \pm 0.9) \times 10^{-9}. \ee 
In this model, both of the mass eigenstates contribute to \gm and one can estimate the contribution using expressions given in \cite{fajfer}. Keeping $M_1 = 1100$ GeV, the leptoquark contribution to \gm is given as,
\begin{align}
a_\mu^\Delta &= 1.34 \times 10^{-6} ~ \lambda_2 \lambda_3 - \frac{10^{-9}}{(M_2/\text{GeV})^2} \left( 6.11 \lambda_1^2 + 5.53 \lambda_2^2 - 9.4 \times 10^4 \lambda_2 \lambda_3 + 5.53 \lambda_3^2\right) + ... \\
&\approx 1.34 \times 10^{-6} ~ \lambda_2 \lambda_3  - 10^{-11} \left( 
8.65~\lambda_1^2 + 7.83 ~\lambda_2^2 + 7.83~\lambda_3^2
\right) + \mathcal{O}(10^{-13})
\end{align}
where the approximation is obtained using the benchmark point $M_2 = 1000$ GeV. From the above expressions one can see that the leading contribution does not depend on $M_2$. It is also clear that the product $\lambda_2 \lambda_3 \approx 10^{-3}$ gives the correct estimate for \gm. In Section 6 we use $a_\mu^\Delta = \delta a_\mu$ to constrain the parameter space of the model. 

\section{Flavor Anomalies}\label{flavor}
In the last two decades, loop-induced $b \rightarrow s$ transitions have been playing an active role in understanding the physics beyond the Standard Model. Starting from the first observation of $B\rightarrow K^* \gamma$, many decays involving $b\rightarrow s$ transitions have been observed. Two of the key observables for LFU violating decays of the B meson are $R_K$ and $R_{K^*}$, defined as 
\begin{equation}
R_{K^{(*)}}=\frac{\mathcal{BR}(B\rightarrow K^{(*)}\mu\mu)_{q^2\in[q_1^2,q_2^2]}}{\mathcal{BR}(B\rightarrow K^{(*)}ee)_{q^2\in[q_1^2,q_2^2]}}.
\end{equation}
It was shown in \cite{HillerOriginal} that within the SM, the hadronic uncertainties in these expressions cancel which results in \rk,\rks $\approx$ 1. However recent measurement of \rks~by LHCb has reported $2.1-2.3 \sigma$ and $2.3-2.5\sigma$ deviations in the low-$q^2$ (0.045 - 1.1 GeV$^2$) and central-$q^2$ (1.1 - 6 GeV$^2$) regions, respectively \cite{rkstar}. A deviation of $2.6 \sigma$ from SM has also been reported in $R_K$ \cite{rk}. We use the standard prescription of effective Hamiltonian to evaluate the contribution of the Leptoquark to \rk~and \rks.\\

\noindent The most general effective Hamiltonian for $b \rightarrow s l^-l^+$ is given as
\begin{equation}
\label{hamiltonian}
\mathcal{H}_{eff}=-\frac{4 G_f}{\sqrt{2}} V_{tb}V_{ts}^* \left[\sum_{i=1}^6 \mathcal{C}_{i}\mathcal{O}_i +\sum_{i=7}^{T5}(\mathcal{C}_i\mathcal{O}_i+\mathcal{C}^\prime_i\mathcal{O}^\prime_i) \right]
\end{equation}
where $\mathcal{O}_i$ are the operators and $\mathcal{C}_i$ are the Wilson Coefficients (WCs) which can be written as 
\be \mathcal{C}_i=\mathcal{C}_{i}^{SM}+\delta\mathcal{C}_i \ee 
where $\delta \mathcal{C}_i$ represent the shifts due to new physics. Global analyses have been performed to fit $\delta \mathcal{C}_i$ to the experimental results which yield interesting correlations between various WCs \cite{crivellin1, bardhannew}. The operators relevant for the model are
\begin{equation}\label{operator}
\mathcal{O}_9  =\frac{e^2}{(4 \pi)^2} (\bar{s}\gamma_{\mu}P_L b)(\bar{\mu}\gamma^{\mu}\mu), \quad \text{and} \quad \mathcal{O}_{10}  =\frac{e^2}{(4 \pi)^2} (\bar{s}\gamma_{\mu}P_L b)(\bar{\mu}\gamma^{\mu} \gamma_5 \mu).
\end{equation}
The expressions for all other operators can be found in \cite{operators}. As usual, the doubly CKM suppressed contributions from $V_{ub}V^*_{us}$ have been neglected. \\ 

\begin{figure}[H]
	\centering
	\includegraphics[height=3.5cm, width=7cm]{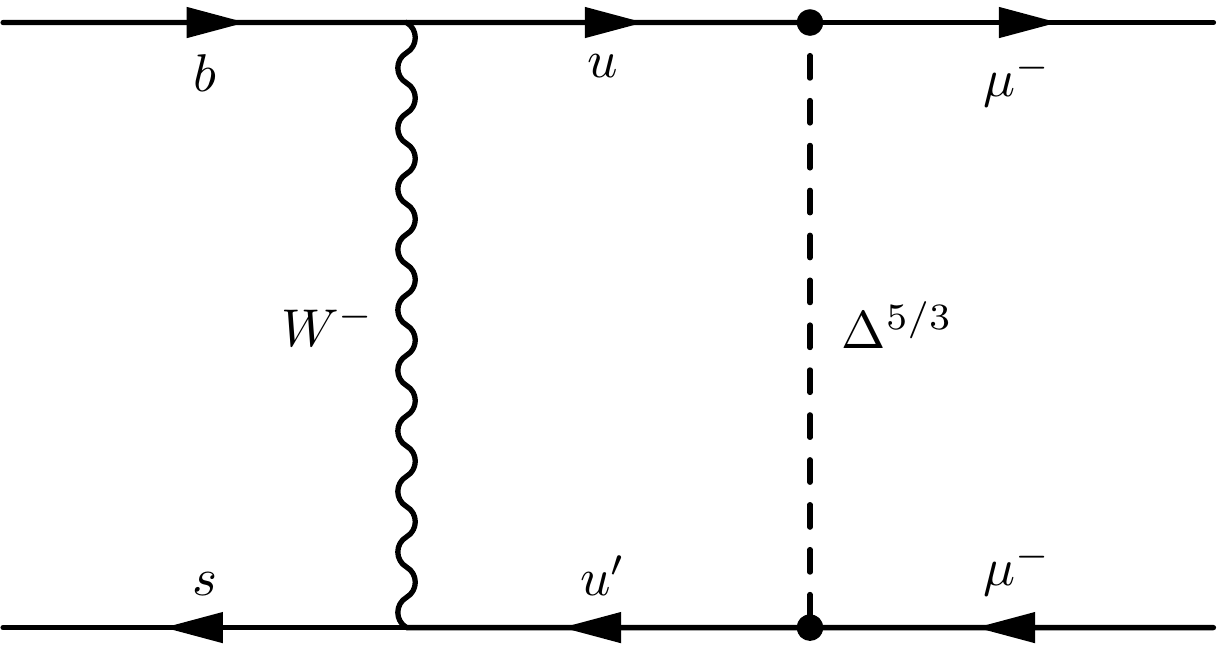}
	\caption{\label{boxdiagram} The box diagram contributing to $b\rightarrow s \mu^-\mu^+$}
\end{figure}

\noindent For the model in consideration, the Leptoquark contributes to $b \rightarrow s \mu^+ \mu ^-$ at one-loop level (Fig. \ref{boxdiagram}) and results in non-zero $\delta \mathcal{C}_9$ and $\delta \mathcal{C}_{10}$ only.
\noindent Using $x_i = (m_i/m_W)^2$, we can write, 
\be
\delta \mathcal{C}_9 = A_1 + A_2 \quad \text{and} \quad \delta \mathcal{C}_{10} = -A_1 + A_2
\ee
where, 
\begin{align}
A_1 &= \frac{\left|\lambda_2\right|^2}{8 \pi \alpha_{em}}\mathcal{F}_1(x_t,x_{t}), \\ 
A_2 &= -\sum_{u,u'\in c,t} (Vy_R)_{u\mu}^\star(Vy_R)_{u'\mu}\frac{1}{16 \pi \alpha_{em}} \frac{V_{ub}V_{u's}^*}{V_{tb}V_{ts}^*}\mathcal{F}_2(x_u,x_{u'}) , \\
\mathcal{F}_1(x_u,x_{u'}) &=\frac{\sqrt{x_ux_{u'}}}{4} \left[\frac{x_{u'}(x_{u'}-4)\log x_{u'}}{(x_{u'}-1)(x_{u}-x_{u'})(x_{u'}-x_{\Delta})} \right. \\ 
&+ \frac{x_{u}(x_{u}-4)\log x_{u}}{(x_{u}-1)(x_{u'}-x_{u})(x_{u}-x_{\Delta})} 
\left. -\frac{x_{\Delta}(x_{\Delta}-4)\log x_{\Delta}}{(x_{\Delta}-1)(x_{\Delta}-x_{u'})(x_{\Delta}-x_{u})} \right], \nonumber \\
\mathcal{F}_{2}(x_u,x_{u'}) &= \frac{x_u^2 \log x_u}{(x_u-x_{u'})(x_u-x_{\Delta})}+ \frac{x_{\Delta}(x_u+x_{u'}-x_ux_{u'})\log x_{\Delta}}{(x_u-x_{\Delta})(x_{\Delta}-x_{u'})}\\ &+\left[\frac{x_u^2-1}{(x_u-x_{\Delta})(x_u-x_{u'})}+\frac{x_{u'}^2}{(x_{u'}-x_{\Delta})(x_{u'}-x_u)}\right]\log x_{u'}. \nonumber
\end{align}
The contribution of up-quark is CKM suppressed. We have used \emph{Package-X} \cite{packagex} and the unitary gauge to evaluate the loop-functions $\mathcal{F}_1$ and $\mathcal{F}_2$. \\

\noindent To evaluate $R_{K}$ and $R_{K^{\star}}$ from the WCs, we use the simplified expressions from \cite{rkdefinition} and obtain, 
\be
\label{rk1} 
R_K= 1. + 0.49 A_1+0.06 A_1^2-0.01 A_2+0.06 A_2^2 
\ee
\be
\label{rks1} R_{K^{\star}}=1. +0.47A_1+0.07 A_1^2-0.14 A_2+0.07 A_2^2.
\ee
Immediately one can observe that the solution $-1<A_1<0$ and $A_2 = 0$ is consistent with latest results. This was also the conclusion in \cite{becirevic}. \\

\noindent Recent measurement $B_s\rightarrow \mu^-\mu^+$ by LHCb is in close agreement with the SM and provides a constraint on the model \cite{lhcb317}.  In the operator basis \eqref{hamiltonian}, branching ratio of $B_s\rightarrow \mu^-\mu^+$ can be written as \cite{becirevicnew}
\begin{equation}\label{Bs}
\begin{split}
\mathcal{BR}(B_s\rightarrow \mu^-\mu^+)= \frac{\tau_{B_s}}{16 \pi^3} \frac{\alpha^2 G_F^2}{m_{B_s}^3}f_{B_s}^2\left|V_{tb}V_{ts}^{\star}\right|^2 m_{B_s}^6 m_{\mu}^2 (1-\frac{2 m_{\mu}^2}{m_{B_s}^2}) \left|\mathcal{C}_{10} \right|^2
\end{split}
\end{equation}
In general, this process gets contribution from  $\mathcal{C}^{\prime}_{10},~\mathcal{C}^{(\prime)}_S$ and $\mathcal{C}^{(\prime)}_P$ as well.  However, we are ignoring them as these WCs are zero in the SM as well as the model under consideration. In the SM, $\mathcal{B}(B_s\rightarrow \mu^+\mu^-)$ is $(3.65\pm 0.23)\times 10^{-9}$  \cite{bobeth} while LHCb has measured it to be $2.8^{+0.7}_{-0.6}\times 10^{-9}$ \cite{lhcb317}. For the model considered in this paper, \eqref{Bs} is 
\begin{equation}
\label{bsmumu}
\mathcal{BR}(B_s\rightarrow \mu^-\mu^+)= 10^{-9}(3.4+1.65(A_1-A_2)+0.2 (A_1-A_2)^2)
\end{equation}
using parameters given in \cite{becirevicnew}. Again, one can see that the solution $-1<A_1<0$ and $A_2 = 0$ is consistent with the experiments. With these expressions, one can write the observables in terms of the couplings as, 
\be
\label{num_rk}
\begin{split}
	\text\rk = 1. -\left(5.16\times10^{-2}\right) \text{$\lambda_2$}^2 +  \left(6.66\times 10^{-4}\right) \text{$\lambda_2$}^4 \\ 
	-\left(1.66\times 10^{-5}\right) \text{$\lambda_3$}^2+\left(1.59\times 10^{-7}\right)\text{$\lambda_3$}^4 
\end{split}
\ee

\be
\label{num_rks}
\begin{split}
	\text\rks = 1. -\left(4.96\times  10^{-2}\right) \text{$\lambda_2 $}^2 +  \left(8.18\times 10^{-4}\right) \text{$\lambda_2 $}^4\\
	-\left(2.34\times 10^{-4}\right) \text{$\lambda_3$}^2 +\left(1.96\times 10^{-7}\right) \text{$\lambda_3 $}^4
\end{split}
\ee

\be
\label{num_bs}
\mathcal{BR}(B_s\rightarrow \mu^-\mu^+) = 2.01 \times 10^{-10} \left| 4.1 - 0.10~\lambda_2^2 - 1.6\times10^{-3}~\lambda_3^2 \right|^2
\ee
In passing, one can note that these expressions do not explicitly depend on $\lambda_1$. This is due to the fact that the term proportional to $\lambda_1$ will enter the expression due to u-quark in the loop which is CKM suppressed. Henceforth, the term 'flavor anomalies' will be used to refer to \rk~and \rks~with imposed constraints from $\mathcal{BR}(B_s\rightarrow \mu^-\mu^+)$.
\section{IceCube PeV Events}\label{IC}

During the first four years of its operation, the IceCube neutrino observatory at the South pole has observed more number of PeV events than expected. This has resulted in a lot of interesting studies in various fields \cite{kohri1, kohri2, atri}. Resonant production of \lq by interactions of astrophysical neutrinos with partons has been proposed as a possible explanation of the excess in PeV events at IceCube \cite{iclq1, iclq2,iclq3,iclq4,iclq5,bhupalice}. In the model considered in this paper, the following neutrino interactions are possible: 
\begin{align*}
&\text{Neutral Current (NC) Like: \hspace{1cm} }\bar{\nu}_i u~ \overset{\da}{\longrightarrow} \bar{\nu}_j u;~\bar{\nu}_j  t \hspace{1cm} {i,j = e,\mu,\tau} \\ 
&\text{Charged Current (CC) Like: \hspace{1cm}} \bar{\nu}_i u~ \overset{\da}{\longrightarrow} \mu d;~ \mu b  \hspace{1cm} {i = e,\mu,\tau} \\
\end{align*}
 It is important to distinguish between the CC and NC interactions due the difference in their deposited energy signature \cite{olga, gandhiuhe}. Ideally speaking, one should also distinguish between shower and track events as the observed PeV events are only shower type. However, one can attribute this to the smallness of statistics and hence we do not consider this difference.  \\ 

\noindent The number of events due to \lq contribution in the deposited energy interval $\left(E_i, E_f\right)$ is \cite{iclq4, olga}
\begin{equation}
\label{olgarate}
\mathcal{N} =T ~N_A~ \int_{0}^{1}dy \int_{E_{\nu}^{ch}(E_i, y)}^{E_{\nu}^{ch}(E_f, y)}dE_\nu ~ \mathcal{V}_{eff}(E_{dep}^{ch})~ \Omega(E_\nu)~ \dfrac{d\phi}{dE_\nu} \dfrac{d\sigma}{dy}^{ch}
\end{equation}
where $T = 1347$ days is the total exposure time, $N_A = 6.023\times 10^{23}$ cm$^{-3}~$ water equivalent is the Avogadro's Number, and $ch$ denotes the interaction channel (NC or CC). Other terms in the expression are discussed in \cite{olga}. For each neutrino or anti-neutrino flavor, an isotropic, power-law flux parametrized as 
	\begin{equation}
	\label{flux}
	\dfrac{d \Phi}{dE_\nu} = \phi_0 \left(\frac{E_\nu}{100~\text{TeV}} \right)^\gamma
	\end{equation}
is assumed.	The best fit values from IceCube \cite{aartsen}
	\be \phi_0 = (2.2 \pm 0.7) \times 10^{-8} \text{GeV}^{-1} s^{-1} sr^{-1} cm^{-2} \ee
	 \be \gamma = - 2.58 \pm 0.25\ee
are obtained using likelihood analysis of the data from 10 TeV - 10 PeV. We use the central values in our analysis. \\
	
\noindent It is evident from the structure of coupling matrices \eqref{coup} that the model only admits interactions between incoming antineutrino (neutrino) with u- and t- (anti-u- and anti-t-) quarks.  It is seen that the Parton Distribution Function (PDF) of t-quark is negligible as compared to that of u-quark. Hence, we only consider interaction with u-quark in our analysis. The differential cross-section for this process is given as \cite{iclq4}
	\be
	\label{crosssection}
	\dfrac{d\sigma}{dy}^{NC/CC}=\frac{\pi}{2}~ \frac{\Lambda_{NC/CC}^4}{|\Lambda^2| }~ \frac{\mathcal{U}(M_\Delta^2/s,yM_\Delta^2)}{s}
	\ee
	where $s = 2 M_N E_\nu$, and $\mathcal{U}(x,Q^2)$ is the PDF of u-quark in an isoscalar proton evaluated at energy $Q^2$. In terms of the \emph{valence} and \emph{sea} quark distributions, one can write \cite{iclq1}
	\be
	\mathcal{U} = \frac{u_{v+s} + d_{v+s}}{2}.
	\ee
	We have used the Mathematica package MSTW \cite{mstw} to obtain these PDFs. \\
	
\noindent The dependence of event rate on couplings is captured by
	\be
	\Lambda_{NC}^4 = \lambda_1^2 \left( \lambda_1^2 + \lambda_2^2\right)
	\ee
	\be
	\Lambda_{CC}^4 = \lambda_1^2 \left(\lambda_3^2\right)
	\ee
	\be
	\Lambda^2 =  \lambda_1^2 + \lambda_2^2 + \lambda_3^2
	\ee 	
Given the mass of the \lq ($M_2$) and the couplings, we are now in a position to estimate the contribution of \lq to the IceCube HESE events. We use the standard $\chi^2$ analysis to estimate the couplings that provide the best fit to the data. In order to estimate whether adding \lq contribution results in a better or worse fit to data, we use the statistic
\be
\delta \left( \lambda_i^2, M_{LQ} \right) = 100 \times \frac{\chi^2_{SM} - \chi^2_{SM + LQ}}{\chi^2_{SM}}
\ee
which represents the percentage change in $\chi^2$. We only use the data for which non-zero number of events are observed at IceCube.

\section{A simultaneous explanation}\label{results}
\begin{figure}[H]
	\centering
	\includegraphics[ width = 7 cm]{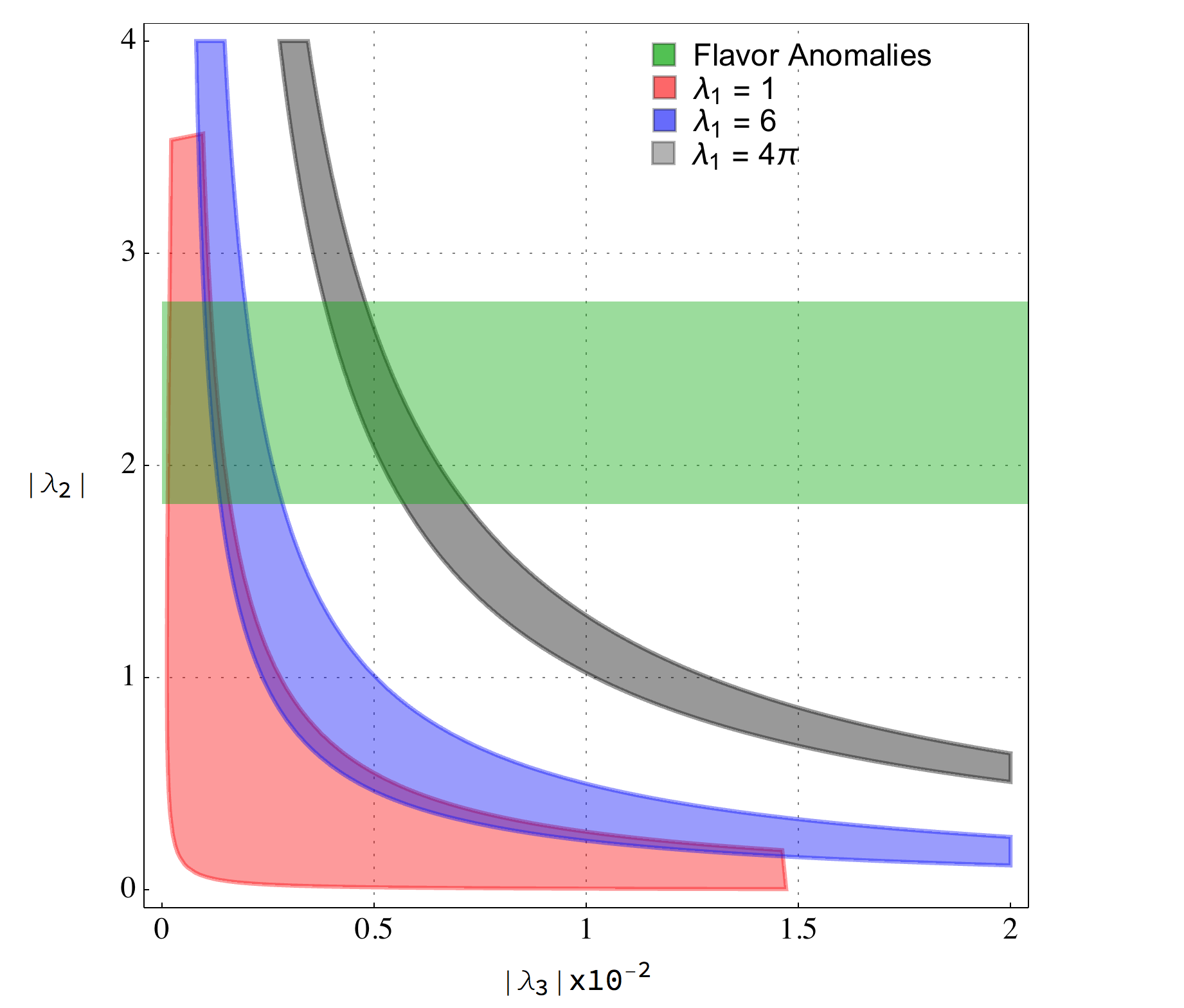}
	\caption{\label{fig:gmflav} The parameter space of \gm  various choice of coupling $\lambda_1$ is shown along with the constraints from flavor anomalies for $M_1 = 1100$ GeV and $M_2 = 1000$ GeV.}
\end{figure}

In this model, we have four free parameters as was pointed out before. However, the Leptoquarks state $\da$ does not feature in any explanation of the flavor anomalies and hence these do not depend on $M_2$. It is also seen that for $M_2 \in (600-1400)$ GeV, the dependence of \gm~on $M_2$ is very weak. Hence, the flavor anomalies and \gm~effectively depend only on the three free couplings in the model. In Fig. \ref{fig:gmflav}, we have shown the parameter space that explains the flavor anomalies a \gm~for $M_1 = 1100$ GeV and $M_2 = 1000$ GeV. \\

\begin{figure}[H]
	\centering
	\includegraphics[height = 5 cm, width = 7 cm]{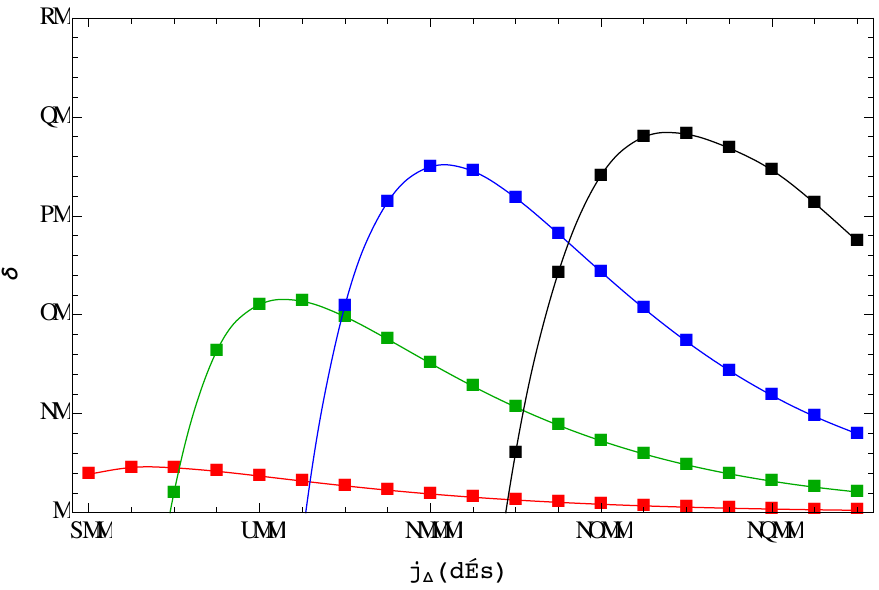}
	\caption{\label{fig:percent} The variation of $\delta$ with $M_2$ for various choice of coupling $\lambda_1$ is shown. The red, green, blue, and black lines correspond to $\lambda_1 = ~1, ~3, ~6, \text{and~} 4\pi$ respectively. }
\end{figure}

\begin{figure}[H]
	\centering
	\includegraphics[width = 10 cm]{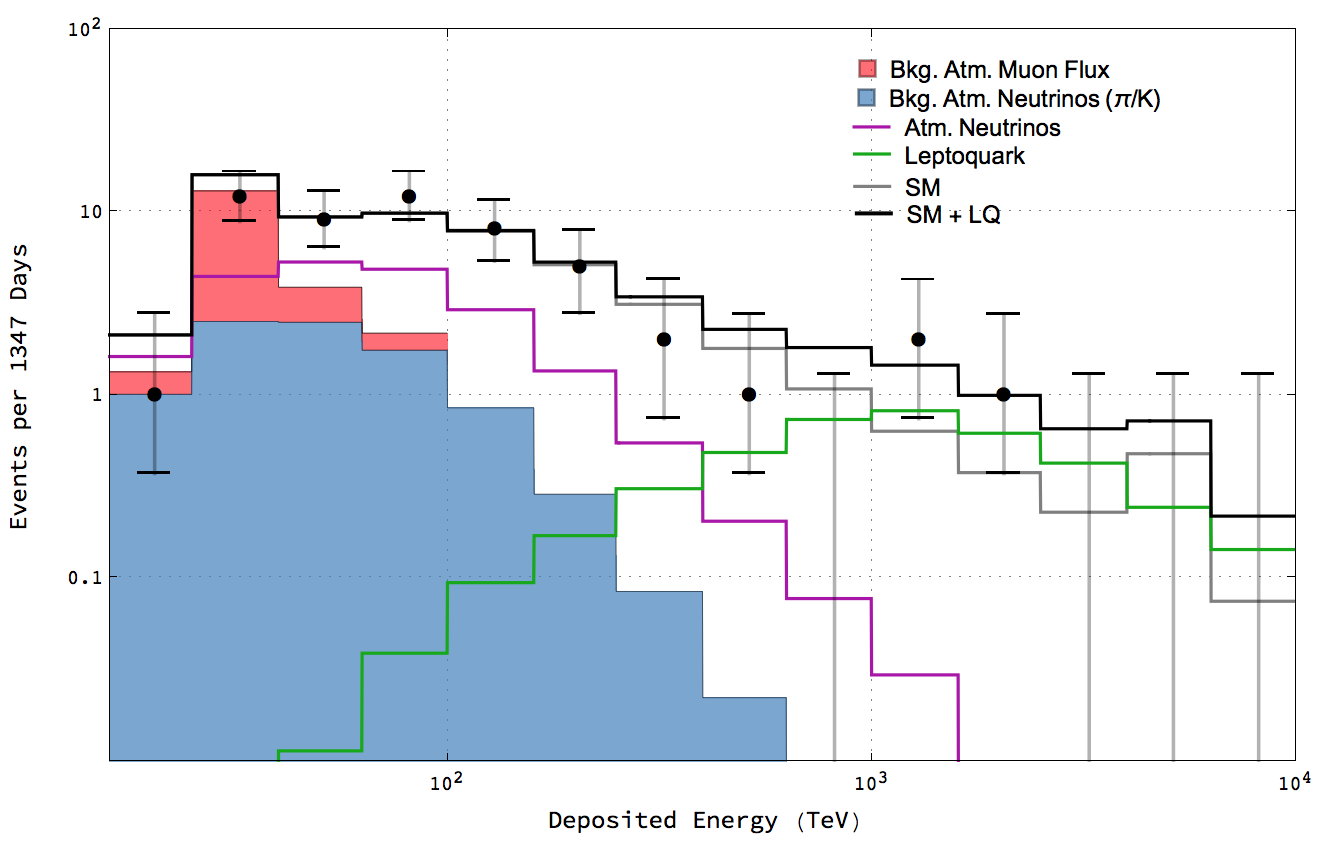}
	\caption{\label{fig:ice} The solid black line shows the prediction for IceCube using Leptoquark and SM interactions.}
\end{figure}

\noindent It can be seen from Fig. \ref{fig:gmflav} that the resolution to flavor anomalies requires $ \lambda_2 \sim \mathcal{O}(1) $ whereas \gm~constrains $\lambda_3 \sim \mathcal{O}(10^{-3})$ for $\lambda_1 < 6$. Using this, and equations (32)-(36), one sees that the number of events at IceCube only depends on the coupling $\lambda_1$. Since, $\db$ does not feature in the explanation for IceCube, these predictions are independent of $M_1$ and only depend on $M_2$. In Fig. \ref{fig:percent}, we show the variation of the statistic $\delta$ with $M_2$ for various choice of coupling $\lambda_1$.  It can be seen that a \lq of mass 800 - 1400 GeV can give 20-35\% improvement to the fit. In Fig. \ref{fig:ice}, we show the contribution of \lq for the benchmark point $M_{LQ} = $1 TeV, $\lambda_1 \approx 6.$ which gives $\delta \simeq 35$. \\

\noindent It is evident that for the aforementioned choices of \lq parameters, one can satisfactorily explain the observed excess in the IceCube HESE Data. However, such an explanation requires large couplings and TeV scale Leptoquarks. Such a scenario should be testable at LHC and is the subject of study in the next section.

\section{LHC constraints}
\begin{figure}[h]
	\centering
	\includegraphics[width=7cm]{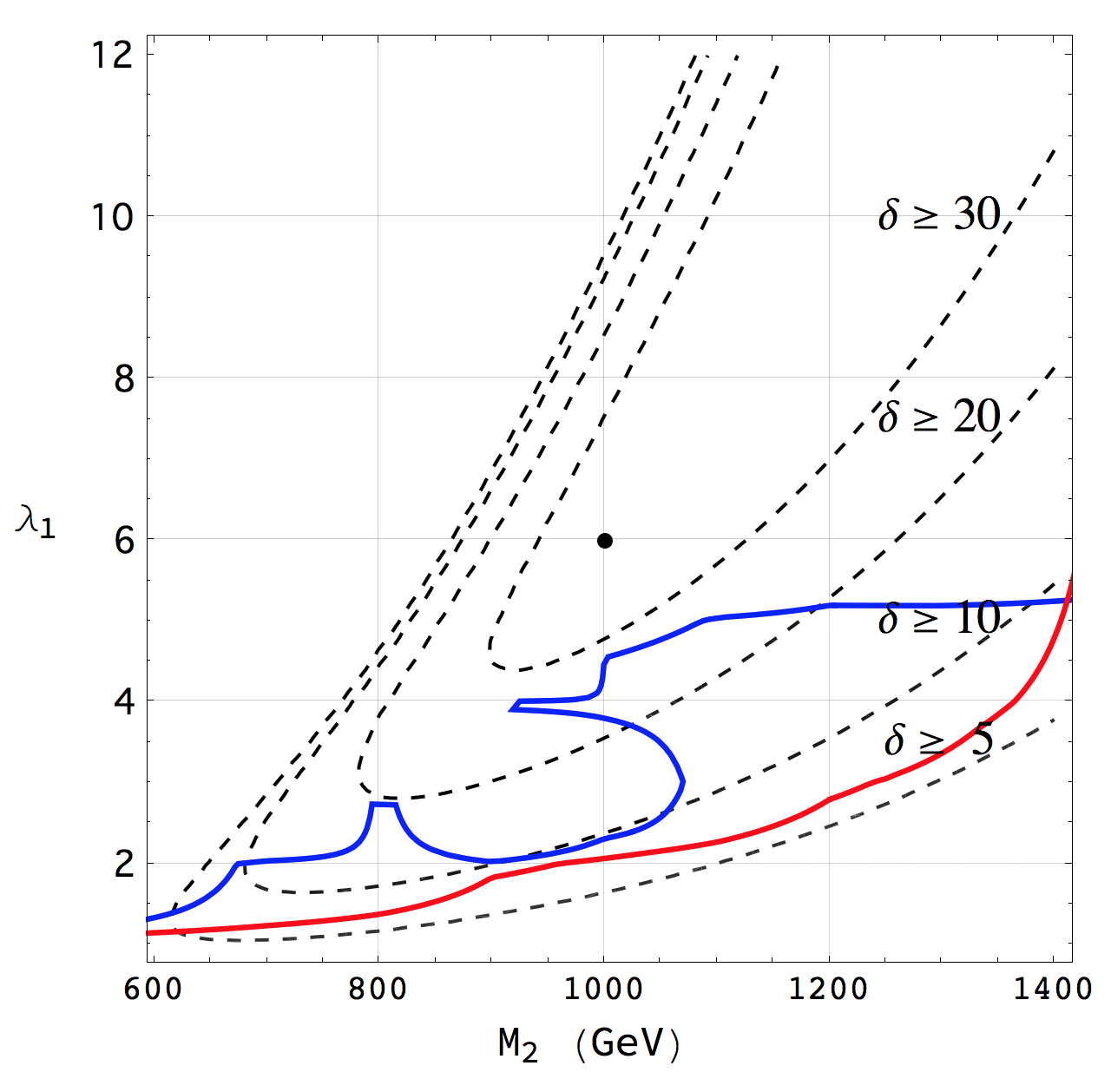}
	\caption{\label{fig:lhc} The Dijet constraints are shown in Blue and the Monojet constraints are shown in Red. The parameter space above the curves is ruled out. The contours of $\delta$ are shown and the benchmark point used to generate Fig. \ref{fig:ice} is shown.}
\end{figure}

Since Leptoquarks carry color charge, they can by singly or pair produced in pp collisions. Subsequent decays of these \lqs in the detector will give rise to jets, leptons, and neutrinos. This gives very interesting final states of the form $jjll,~jjl\nu,~jl\nu,~jj\nu,~j\nu\nu$, etc and has been the subject of various studies \cite{lqlhc1,lqlhc2,lqlhc3,lqlhc4,lqlhc5,lqlhc6,lqlhc7,lqlhc8,lqlhc9,lqlhc10,lqlhc11}. As these neutrinos are not seen by the detector, they appear as a Missing Transverse Energy (MET). For the LHC analysis, we have implemented the model using FeynRules (v2) \cite{feynrules} and simulate the above processes using MadGraph (v5)\cite{madgraph} which uses Pythia (v8) \cite{pythia} for parton showering. We then use CheckMATE (v2) \cite{checkmate} to find the value of statistical parameter, r defined as 
\begin{equation}
r=\frac{(S-1.96\Delta S)}{S_{exp}^{0.95}}
\end{equation} for several points in the parameter space. Here, $S$ and $\Delta S$ represents signal and its uncertainty. The numerator represents $95\%$ confidence limit on number of events obtained using CheckMATE and the denominator represents $95\%$ experimental limits on the number of events. The approximate functional form is obtained using linear interpolation. Parameter space with $r\geq1$ is excluded and the results are summarized in Fig. \ref{fig:lhc}. \\

\textbf{Constraints from $jjll$:} When the Leptoquarks are pair produced in pp collisions, each Leptoquark can decay into a charged lepton and a quark. Recently, ATLAS collaboration performed a search for new physics signature of lepton-jet resonances based on $\sqrt{s}=13$ TeV data \cite{dijetdilep} wherein pair production of Leptoquarks was studied based on events like $eejj$ and $\mu\mu jj$. The analysis gives an upper limit on branching ratio of first and second generation Leptoquark to $ej$ and $\mu j$ respectively. Although, our model has inter-generation couplings, we use these limits to constrain the free parameters in our model. We find that, 
\begin{equation} \mathcal{BR}\left( \lqa\to \mu j \right) \approx 1 \end{equation}
as it couples to only second generation of leptons. This puts a lower limit on mass of Leptoquark as,
$$M_{1} \geq 1100~\text{TeV}.$$
We use the lower limit to generate other constraints and for flavor analysis. For $\Delta^{2/3}$ state, 
\begin{equation}
\mathcal{BR}\left( \lqb\to \mu j \right) \propto \lambda_4^2 \approx 0
\end{equation}  
which does not provide any constraints from this analysis. \\

\textbf{Constraints from $jj\nu\nu$:} When the Leptoquark state $\Delta^{2/3}$ is pair produced, each can decay into a neutrino and a quark giving rise to a peculiar Dijet + MET signature. The parameters $M_1$ and $\lambda_2$ are fixed from flavor observables and this process only depends on $M_2$ and $\lambda_1$. We use the 13 TeV ATLAS search \cite{dijetmet} to find constraints on this parameter space. \\

\textbf{Constraints from $j\nu\nu$:} If the Leptoquark $\Delta^{2/3}$ is singly produced, it can decay into a quark and a neutrino giving rise to Monojet signal at the LHC. Again, this process only depends on the parameters $M_2$ and $\lambda_1$. We use the 8 TeV ATLAS search \cite{monojet} to find constraints on this parameter space. \\

\textbf{Other Constraints:} We find that the Monojet constraints are strong enough to rule out the entire parameter space that explains IceCube PeV events and we do not provide results for other processes. However, in passing, we note that the constraints from $jl\nu$ final state are much stronger. This maybe relevant for future tests of Leptoquark models.  \\

\section{Conclusion}

The discrepancy in anomalous magnetic moment of muon, the observed excess in PeV events at IceCube, and the lepton flavor universality violation in B decays are some of the biggest challenges facing the Standard Model. A simultaneous explanation for these problems is desirable. An ad-hoc solution such as Leptoquarks, if it can successfully address these issues, will shed more light on the unification scenarios that contain them. One such attempt was made in this paper using a scalar doublet Leptoquark. The peculiar feature of this model is that the flavor anomalies are explained at one-loop level. Because of the loop suppression, one does not require either very small couplings or very heavy Leptoquarks. We find that one can explain the B-anomalies \rk~and \rks~with $\mathcal{O}(1)$ coupling and TeV scale Leptoquark. In the past, similar parameters have been invoked to explain IceCube events and a unified explanation seemed possible. However, we find that in order to explain IceCube data, one \emph{needs} \lq coupling to first generation quarks and neutrinos. This coupling will give rise to Monojet and Dijet signals at LHC, both of which are severely constrained. Because of this, any attempt to explain IceCube events using such Leptoquarks would be in conflict with LHC data. This conclusion was also reached for a Scalar Triplet in \cite{iclq1}, and for Scalar Singlet in \cite{lqlhc8}. Any unification scenario that has \lq like states, IceCube explanation in such theories (e.g. R-Parity Violating MSSM \cite{bhupalice}) should also be in conflict. \\

\section*{Acknowledgements}
The authors would like to thank Prof. Namit Mahajan and Prof. Subhendra Mohanty for invaluable discussions and suggestions. The authors also thank the anonymous referee for pointing out the important LHC constraints on the model.

\bibliographystyle{unsrt}
\bibliography{references_dg12082.bib}

\end{document}